**Small antimicrobial resistance proteins (SARPs) – Small proteins conferring antimicrobial resistance**

**Authorship:** Rianne C. Prins, Sonja Billerbeck*

**#Affiliations**

Molecular Microbiology, Groningen Biomolecular Sciences and Biotechnology institute, University of Groningen, Groningen, The Netherlands

**\*Correspondence:** s.k.billerbeck@rug.nl (S. Billerbeck)

**Abstract**

Small open reading frames are understudied as they have been historically excluded from genome annotations. However, evidence for the functional significance of small proteins in various cellular processes accumulates. Proteins with less than 70 residues can also confer resistance to antimicrobial compounds, including intracellularly-acting protein toxins, membrane-acting antimicrobial peptides and various small-molecule antibiotics. Such herein coined Small Antimicrobial Resistance Proteins (SARPs) have emerged on evolutionary timescales or can be enriched from protein libraries using laboratory evolution. Our review consolidates existing knowledge on SARPs and highlights recent advancements in proteomics and genomics that reveal pervasive translation of unannotated genetic regions into small proteins that show features of known SARPs. The potential contribution of small proteins to antimicrobial resistance is awaiting exploration.

**Highlights**

- Small proteins have been historically overlooked and are still understudied, but evidence of functionally important small proteins is accumulating across all domains of life.

- The functional space of small proteins includes antimicrobial resistance.

- Pervasive transient transcription and translation from noncoding genetic regions is a potential source of *de novo* small proteins for adaptation to environmental stress, including antimicrobial stress.

- Synthetic gene libraries encoding random small proteins can be used as a tool for identification and characterization of small resistance proteins in growth-based assays.

**Main text**

## 1. The long overlooked functional space of small proteins is starting to be discovered and includes antimicrobial resistance

**Small proteins** (see **Glossary**) – herein defined as polypeptide chains less than 100 amino acids (aa) – have a huge sequence space ($10^{26}$ for a 20 aa peptide) that has been harnessed by nature and by laboratory engineering as a reservoir for bioactive compounds such as hormones, binders [1], antimicrobial compounds [2], genetic regulators [3], or building blocks for nanomaterials [4].

The pool of small proteins within a cell has been historically overlooked due to technical complications in their identification and characterization. As in-frame start and stop codons often occur near one another by chance in genomes, the bioinformatic identification of translated and functional small protein coding genes among a vast amount of artifactual short open reading frames (sORFs) proved to be a challenging task [5, 6]. When the genome of the yeast *Saccharomyces cerevisiae* was first sequenced, an *ad hoc* cut-off of 100 codons was therefore used to minimize false-positive predictions of novel open reading frames [7]. Besides from sORFs, small proteins can also be derived from processing of larger proteins such as cleaved signal peptides or proteolytically generated fractions of proteins (**Figure 1A**).

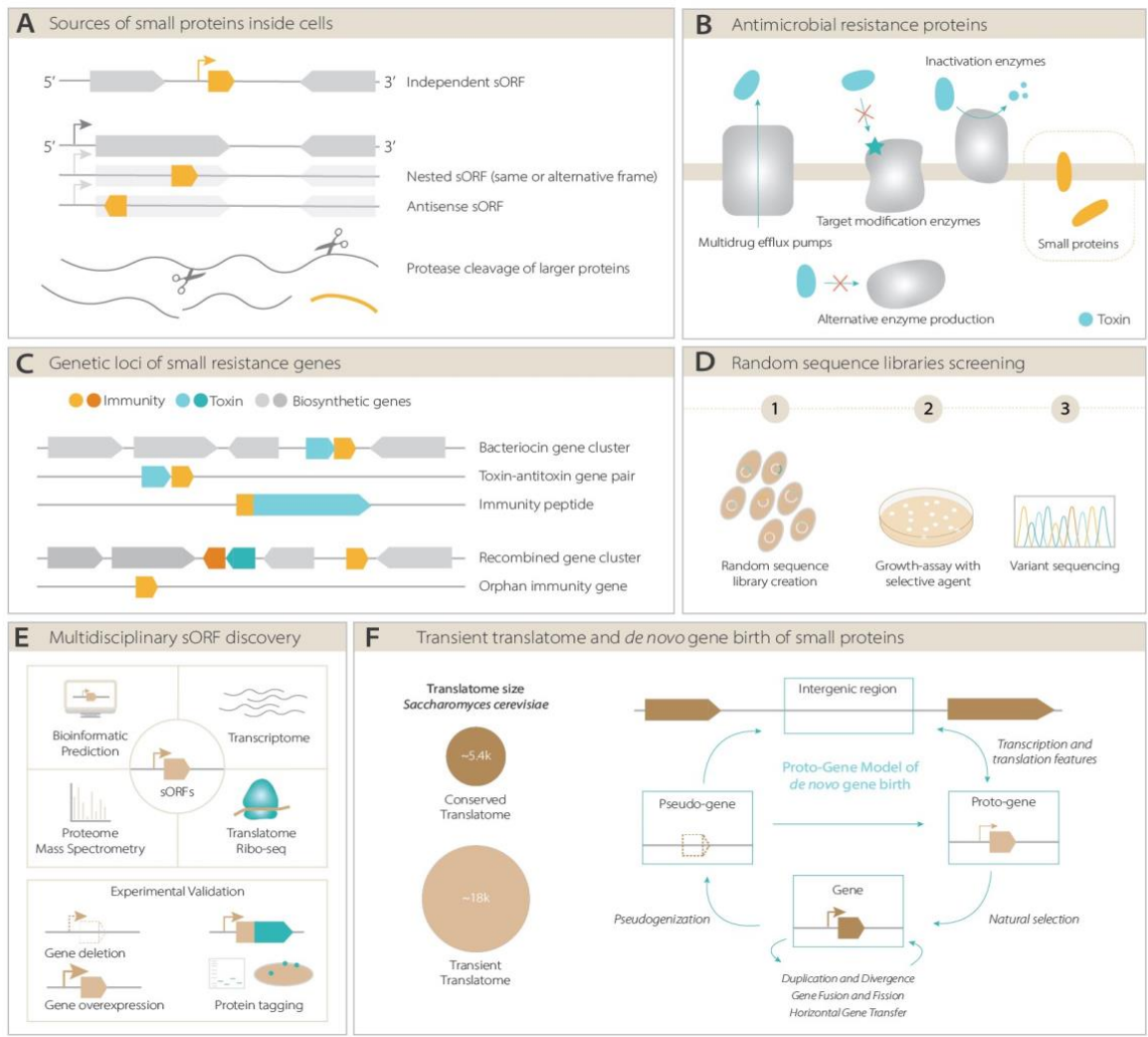

**Figure 1. Small proteins in the context of antimicrobial resistance: (A)** Sources of small proteins inside cells include independent sORFs, sORFs overlapping larger ORFs, and peptides obtained by proteolytic cleavage of larger polypeptides. **(B)** Antimicrobial resistance conferred by small proteins adds to other well-known resistance mechanisms carried out by larger proteins, such as (multi)drug efflux pumps and drug- or target-modifying enzymes. **(C)** Natural small resistance proteins are found in several genomic contexts. Many are found within toxin biosynthetic gene clusters as self-protection agents of producer cells. Other systems lack additional biosynthesis genes, including toxin-antitoxin or some bacteriocin-immunity gene pairs [8]. Small resistance peptides can further be derived from processing of larger protein precursors [9]. Cluster recombination and horizontal gene transfer can lead to clusters that contain multiple small resistance protein (immunity) genes, or orphan immunity genes, allowing a non-producer to become resistant (immunity mimicry) [10–13]. **(D)** Small resistance proteins protective against small molecule antibiotics have been selected from random sequence libraries in the laboratory. **(E)** Novel small proteins are discovered via multidisciplinary approaches, involving bioinformatic predictions, ribosome profiling and proteomics. **(F)** Non-genic genomic regions are pervasively translated and provide a reservoir of transient small proteins that potentially provide fitness benefits [14], and tend to contain predicted transmembrane helices. Non-genic regions can acquire features for transcription and translation. These proto-genes are then subjected to evolutionary selection. Genes that provide fitness benefits may be preserved and evolve towards mature, fixed genes, or degenerate into pseudo-genes if conditions change and selection pressure fades.

Despite the historical cut-off, evidence mounted that sORFs less than 100 codons encode stable, functional proteins. While the first functional small proteins were discovered serendipitously, this inspired innovations in bioinformatic predictions and 'OMICS' technologies for comprehensive identification of sORFs **(Figure 1E)**, leading to increasing numbers of novel potentially functional small proteins [6, 15, 16].

It is now evident that small proteins play vital roles in various cellular processes including cell signaling and communication, membrane transport, cell division, transcriptional regulation and as antimicrobial peptides (reviewed in [17–20]). Furthermore, small proteins are involved in adaptation to changing environments and cellular stress responses [21–24]. For example, Prli42 (31 aa) in *Listeria monocytogenes* is critical for stressosome activation [25] and IroK (21 aa) expression affects tolerance to 3-hydroxypropionic acid in *Escherichia coli* [26].

This leads to one important functional space of small proteins that has gained little attention: Small proteins that render cells resistant to stress imposed by antimicrobial compounds. Only a few small proteins have been mentioned in recent reviews that influence sensitivity to antibiotic stress. For example, in *E. coli* AcrZ (49 aa) modulates the function of the multidrug efflux pump AcrAB-TolC, thereby increasing resistance of the bacterium to a subset of antibiotics [27] and Blr (41 aa), identified in a region that was annotated as 'intergenic', may affect the susceptibility to β-lactams [28, 29].

However, literature contains multiple reports of other small proteins that increase resistance to antimicrobial compounds **(Tables S1-4,** attached at the end of the document). Besides a few reports of small proteins that protect cells against small-molecule antibiotics, the majority consists of self-protection proteins (also called **immunity proteins**) encoded in organisms that produce a protein toxin. In addition, *de novo* small **resistance proteins** have been selected from **random sequence libraries** [30, 31]. Resistance by such small proteins adds to well-known resistance mechanisms conferred by larger proteins, such as (multi)drug efflux pumps or enzymes that modify or degrade the drug, or modify the drug target (**Figure 1B**).

This review aims to consolidate knowledge on small proteins in the context of antimicrobial resistance with emphasis on their mode of protection (**Boxes 1-4**). We also discuss recent advances in proteomics and genomics regarding study of the small proteome of cells (**Figure 1E**), which accentuate its potential to be a source of functional innovation and a potential origin for the *de novo* generation of small resistance proteins (**Figure 1F**).

A specific challenge when exploring the area of small resistance proteins lies in the absence of standardized terminology for this phenomenon. Used nomenclature includes "immunity protein", "antitoxin", "antidote" or "resistance peptide", and small proteins are grouped together with functionally similar larger proteins. We therefore suggest the term "**Small Antimicrobial Resistance Protein (SARP)**" to describe small proteins which confer resistance to antimicrobial compounds, and for the purpose of this review we focus on those consisting of 70 aa or less.

## 2. SARPs that emerged on evolutionary timescales

Microorganisms live in complex microbial communities and secretion of antimicrobial compounds can contribute to a competitive advantage in an ecological niche. The majority of antimicrobial compound-producing microorganisms antagonize closely related species, such that mechanisms of self-protection - often dedicated immunity proteins - are encoded alongside the biosynthesis genes. Besides antibiotic small-molecules, microbes produce an array of protein toxins. As such, there are intracellular **toxin-antitoxin** (TA) **systems,** secreted **bacteriocin-**immunity systems, yeast **killer toxin-**immunity systems, and **contact-dependent effector-**immunity systems. Although these differ in their purpose and mechanism, they all feature a toxin with a corresponding antitoxin/immunity factor that provides protection. Microbial toxins can further be broadly divided into those that act intracellularly and those that primarily target the cell envelope. Several immunity mechanisms are carried out by larger proteins, but others can be the result of rather small proteins (SARPs) [32, 33] as highlighted in the following.

### 2.1. SARPs active against intracellular-acting protein toxins

We identified seven TA systems where the antitoxin is smaller than 70 aa and as such fall within our definition of a SARP **(Table S1-4)**. Most of these small antitoxins neutralize their cognate toxin by formation of a tight intracellular toxin-antitoxin complex **(Box 1)**. Interestingly, the SpoIISABC system in *Bacillus subtilis* contains not one, but two small antitoxins: SpoIISB (56 aa) and SpoIISC (45 aa), that are both able to neutralize the SpoIISA toxin **(Table S2)** [34]. The antitoxins often have a simple, extended conformation with alpha-helices that wrap around the toxin **(Box 1, Figure IA, B, D, E)**. But some form more globular structures **(Box 1, Figure IC)**. Notably, in some cases these antitoxin or toxin-neutralization sequences fused to other protein domains over the course of evolution - specifically to DNA-binding domains that regulate TA system expression [35, 36] **(Box 1, Figure IF).**

Besides intracellular residing TA systems, toxins with intracellular targets can also originate from external sources, including those secreted or injected by neighboring cells. For example, some colicins are contact-independent toxins that hijack specific transport systems to enter target cells, using modular receptor-binding, translocation and cytotoxic domains [37, 38].

There are also contact-dependent toxins, such as the toxin from the TA system VbhAT of

**Box 1: Mode of protection of intracellular small antitoxins within TA systems.**

Bacterial toxins from TA systems disrupt cellular growth by targeting essential cellular

processes. These TA systems are classified based on the nature of the antitoxin, which is

either RNA or a protein [39]. Small protein-antitoxins predominantly belong to Type II TA

modules and neutralize their cognate toxin through tight interactions, which can hinder substrate access to the toxin's catalytic site, induce inactivating conformational changes, or prevent toxin binding to specific targets [40].

Several structures of such toxin-antitoxin complexes have been elucidated (**Figure I**). For

example, the SpoIISA toxin from *Bacillus subtilis* consists of a predicted N-terminal

transmembrane domain and a C-terminal cytoplasmic domain, and a symmetric dimer of the

latter is encased on both sides by a SpoIISB antitoxin (**Figure IE**) [34, 41]. Another structure

demonstrates how the nicked α-helical PaaA2 antitoxin wraps around the ParE2 toxin from *E.*

*coli* (**Figure IA**) [42]. The VbhAT system from *Bartonella schoenbuchensis* and the

*Escherichia coli* FicAT system exhibit similarities and feature an inhibitory helix that

obstructs the catalytic site of the FIC-domain toxin (**Figure IB**) [43, 44]. Although the typical

toxin:antitoxin stoichiometry is 1:1, the VapX antitoxin from *Haemophilus influenzae*

occupies a cleft formed between a VapD toxin dimer (1:2) (**Figure IC**) [45].

The Phd-Doc system showcases a fusion of a small toxin-neutralization domain with a DNA-

binding domain, that dimerizes to repress the TA operon (**Figure IF**) [46]. The neutralization

domain alone suffices for toxin inhibition [36]. These promoter-repressor and antitoxin

domains represent evolutionary separate modules of which recombination contributes to TA

family diversity [47]. Experimental data, verifying the modular neutralization domain, are available for several such antitoxins (**Table S2**), and, in some instances, antitoxins canonically consisting if the two domains are also found in a shorter, neutralization-domain-only format, as exemplified by the *Methanocaldococcus jannaschii* RelBE TA system (**Figure IG**) [35].

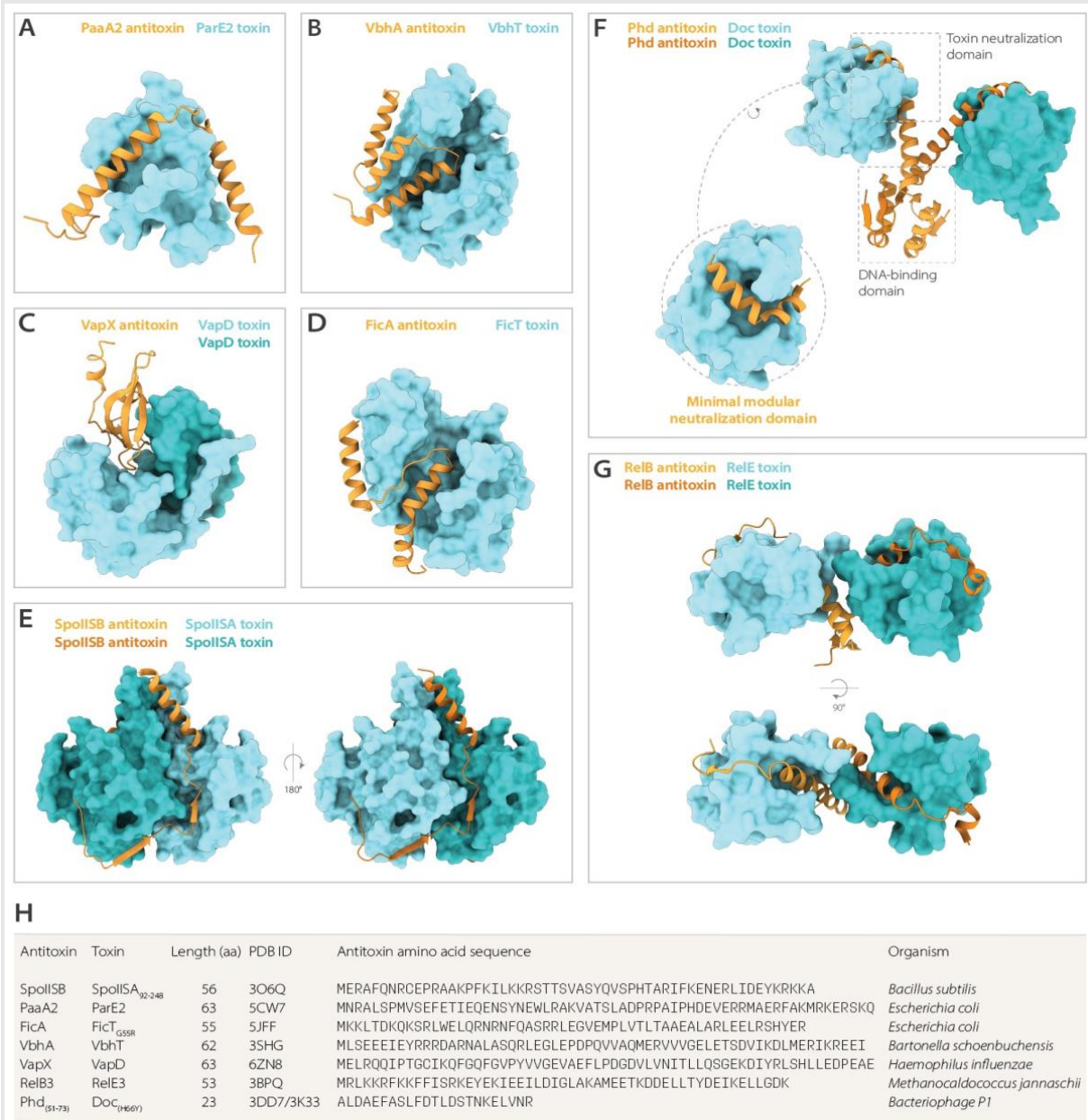

**Figure I. Structures of toxins with small antitoxins.** Structures were retrieved from the Protein Data Bank (PDB). (**A**) PaaA2 and ParE2. (**B**) VbhAT. (**C**) VapXD. (**D**) *Ec*FicAT$_{GSSR}$. (**E**) SpoIISB with C-terminal SpoIISA domain (residues 92-248). (**F**) PhD-Doc (Doc$_{H66Y}$, the small domain formed by Phd residues 51-73 is sufficient for toxin neutralization). (**G**) *Mj*RelBE, consisting of the neutralization domain only. A DNA-binding domain is canonically present similar to that in panel F, but lacking in this natural variant. (**H**) Table overview of properties of displayed proteins.

*Bartonella schoenbuchensis* that is translocated into recipient cells alongside VbhAT-encoding plasmid transfer by a type IV secretion system [10]. This process directly initiates addiction of recipient cells to the incoming plasmid, and recipient cells may benefit from possessing an '**orphan**' chromosomal antitoxin gene copy (of the small antitoxin VbhA, 62 aa) to prevent addiction to invading DNA [10, 48] **(Box 1, Figure IB) (Figure 1C)**.

### *2.2. SARPs active against membrane-acting protein toxins*

We remarkably found a large pool of small resistance proteins – here historically called immunity proteins - that protect against bacteriocins that act on membranes (e.g. via pore formation and dissipation of the membrane potential): Over 30 small proteins, associated with 8 classes of structurally diverse bacteriocins **(Table S1)**. These immunity proteins are all hydrophobic, with one or two predicted transmembrane helices and therefore likely localize to the membrane **(Box 2)**, where they hinder the toxin in ways that remain largely enigmatic **(Box 3)**.

While most of these have been studied in bacteria, one such small hydrophobic immunity peptide has been described in yeast: An immunity peptide (49 aa) that protects *S. cerevisiae* against the membrane-acting yeast killer toxin K2 [9]. In contrast to bacterial immunity proteins that are encoded by sORFs located in the bacteriocin biosynthetic gene cluster, an N-terminal peptide of the K2 precursor, which has features of a signal peptide, provides K2 protection **(Figure 1C)** [9]. This showcases that the phenomenon of such small immunity proteins is not limited to bacteria, and that resistance-conferring small proteins do not only arise from independent sORFs but also from peptides originating from larger proteins, such as processed signal peptides **(Figure 1A)**.

**Box 2. Hydrophobic SARPs against membrane-acting toxins.**

Small resistance proteins – here called immunity proteins - offer protection against a diverse array of class I, II and III gram-positive bacteriocins [49] **(Figure II)**. Class I comprises small (<10 kDa), heat-stable, post-translationally modified bacteriocins with features like lanthionines (lantibiotics), sulphur-to-α-carbon bonds (sactibiotics) or head-to-tail cyclization. Class II peptides are also small but unmodified, including those with a pediocin-like motif or those composed of two peptides. In contrast, class III bacteriocins are larger (>10 kDa) and more heat-labile. Furthermore, similar small immunity proteins were also identified against gram-negative bacteriocins (e.g. microcins) [50], and a protein toxin secreted by yeast (yeast killer toxin K2) [9].

A commonality among these toxins is their ability to target and disrupt membranes, often achieved via pore formation that leads to a disturbed membrane potential. Despite the diverse structures and origins of the toxins, they all share that the cell can protect itself from their action via a structurally relatively simple immunity protein. These immunity proteins typically feature one or two predicted transmembrane helices and are presumed to localize to the membrane - the site of action **(Figure IIA).** Experimental confirmation of membrane localization exists for certain proteins like CbnZ and DysI [33, 51].

The flanking regions around the transmembrane helices often harbor positively charged amino acids, rendering the majority of these small immunity proteins cationic with a high isoelectric point **(Figure IIB).**

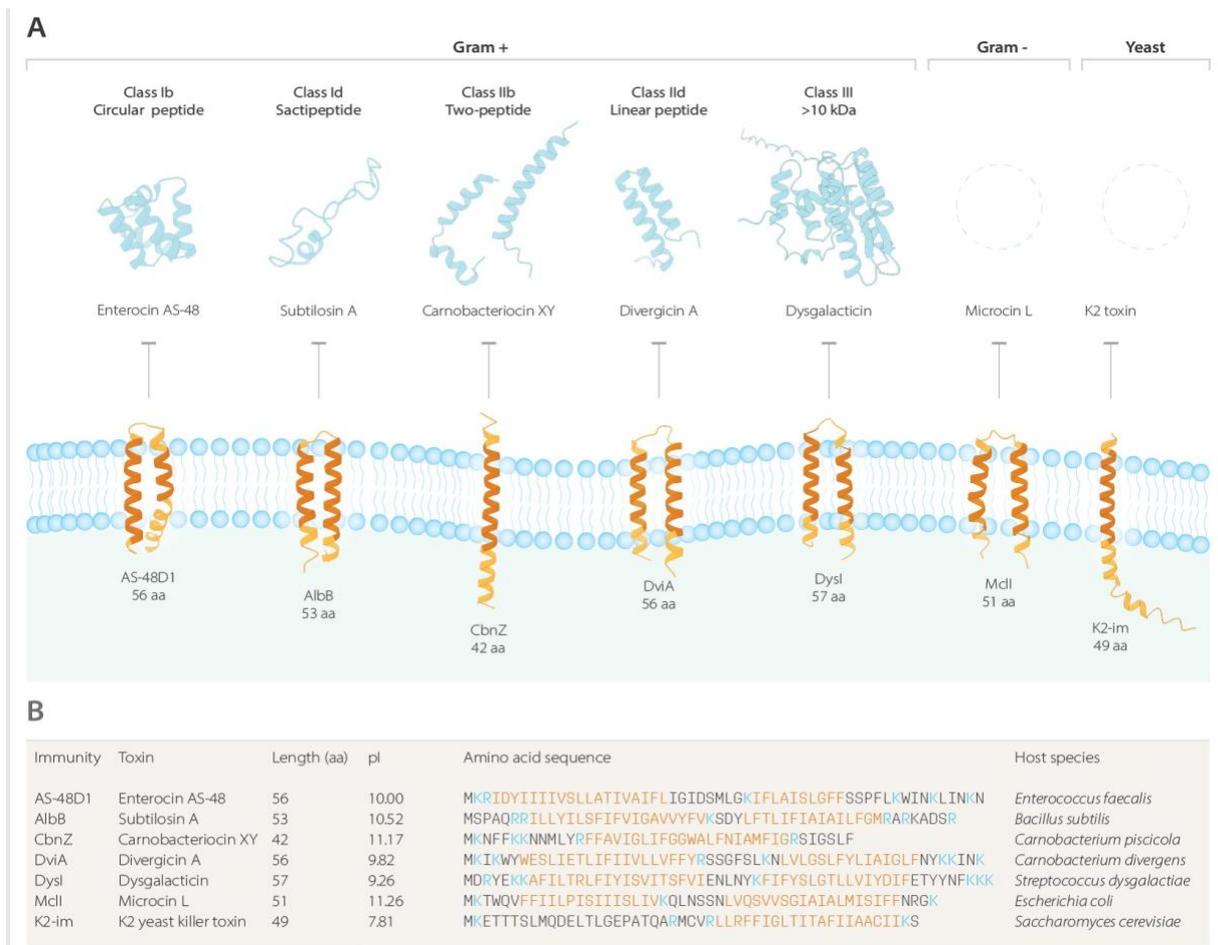

**Figure II. Representative toxin-immunity pairs. (A)** Structures of hydrophobic immunity proteins and cognate toxins from bacteria and yeast: AS-48D1 [52], AlbB [53], CbnZ [33, 54], DviA [8], DysI [51], McII [50], and the K2-im peptide [9]. Membrane directionality was adopted as predicted by DeepTMHMM [55]. The following structures were obtained from PDB: enterocin AS-48 (1O83), subtilosin A (1PXQ), carnobacteriocin XY (5UJR and 5UJQ). Others were predicted using AlphaFold [56]. For microcin L and the K2 toxin no confident structure predictions were obtained. Note that only a single membrane is presented in the figure, whereas yeast and gram-negatives have multiple membrane compartments, and that other cellular factors may be involved in the immunity mechanism (see **Box 3**). **(B)** Overview table of the displayed small immunity proteins (see also **Table S1-4**). The predicted transmembrane regions are colored dark orange and positive charges outside those are colored blue. Besides hydrophobicity, a high isoelectric point (pI) is also a feature of the small bacteriocin-immunity proteins. **Of note**: A deep-transcriptomics study in *S. cerevisiae* indicated that *de novo* proteins have a tendency towards a higher pI with a median of 9.33 (see also **Section 4**) [57].

Membrane-acting toxins with similar cognate small immunity proteins also appear in the context of contact-dependent systems, such as the effector-immunity system Ssp6-Sip6 (67 aa) of *Serratia marcescens*, which is delivered to the target cell via a type VI secretion system [58]. The immunity protein Sip6 has two predicted transmembrane regions.

In addition, an intracellular membrane-targeting TA system has been reported: In *E. coli*, the EcnAB TA system consists of the entericidin B toxin (48 aa-precursor) and the entericidin A antitoxin (41 aa-precursor) [32]. Based on sequence identity (55%) these genes appear paralogous and may have arisen from a gene duplication event. This case illustrates, that small membrane-acting toxins and small membrane-acting antitoxins can be relatively close in sequence space (see also **Section 3** and O**utstanding Questions**).

**Box 3. Mode of protection of membrane-associated SARPs.**

Many secreted protein toxins target membranes. Narrow-spectrum toxins potentially require specific interaction with integral membrane proteins for docking or pore formation (for example the mannose phosphotransferase system [59]), while those exhibiting broad-spectrum activity may directly impact the membrane, as experimentally probed by liposome targeting *in vitro* [60] (**Figure IIIA**). Additionally, receptors might be required at low toxin concentrations but become dispensable at higher toxin concentrations [61].

The mode of protection of dedicated small hydrophobic immunity proteins remains a key unresolved question. Here we discuss several hypotheses that have been proposed for the action of small hydrophobic immunity proteins (**Figure IIIB**): If the toxin requires interaction with a specific membrane protein receptor, the immunity protein might interfere by masking the toxin-binding site on the target [62], or by clogging the formed pore (as shown for SaiA (90 aa), immunity protein to sakacin A [59]). In cases where the toxin directly affects the membrane, the immunity protein may sequester toxin molecules and prevent pore assembly [63]. Some evidence exists for potential direct interactions between certain toxin and immunity proteins, or the formation of multimeric structures by the immunity protein itself [58, 64]. Furthermore, it has been suggested that exposure of positively charged amino acid residues on the outer surface of the cytoplasmic membrane may electrostatically hinder the attachment of positively charged toxins [65]. Alternatively, small transmembrane proteins could trigger an indirect mechanism (**Box 4, Figure IVC**), by association with larger membrane proteins, thereby modulating key cellular processes [24]. Additionally, small proteins that align themselves parallel to the membrane can potentially counteract each other's effect on the biophysical properties of the lipid bilayer itself, a concept that is referred to as the 'reciprocal wedge theory' [32, 66] (**Figure IIIC**).

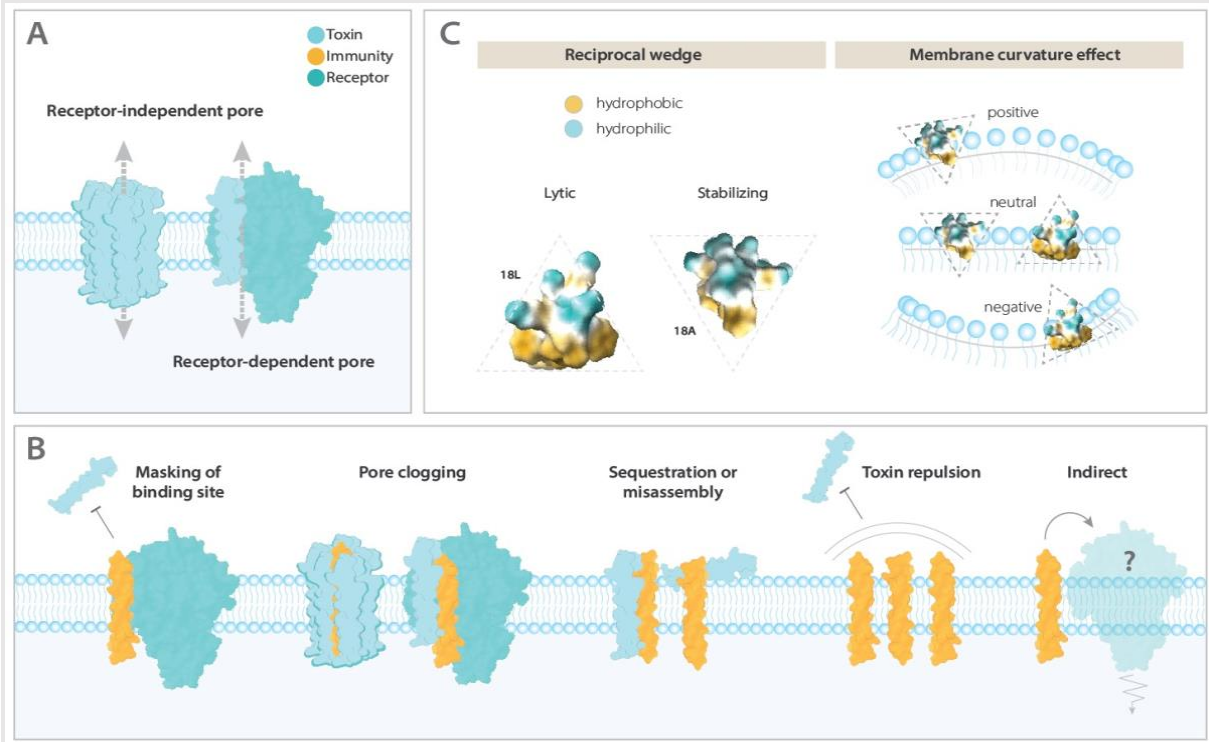

**Figure III. Potential mechanisms of protection at the membrane.**
(**A**) Membrane-targeting toxins typically induce pore formation, potentially *via* a receptor-dependent or -independent way. (**B**) Several mechanisms have been proposed by which small hydrophobic transmembrane proteins could prevent pore formation. (**C**) The reciprocal wedge theory describes how two membrane-associated peptides can have a lytic or stabilizing effect on the cell membrane curvature strain, here exemplified with peptides 18L and 18A [66].

## 2.3. SARPs active against small-molecule antibiotics

There are limited publications demonstrating the natural evolution of small proteins to counter the effects of a small-molecule antimicrobial. One compelling case involves a pentapeptide that enables *E. coli* cells to grow in a four-fold higher concentration of the ribosome-targeting macrolide-antibiotic erythromycin [67, 68]. It came to light following the observation that certain mutations within the 23S ribosomal RNA conferred resistance [69]. Remarkably, upon closer inspection, a sORF was discovered encoding the pentapeptide

'MRMLT', along with a canonical upstream ribosome binding site, which must be actively translated by the ribosome to confer resistance [70–72]. Libraries of random small proteins were used to further characterize the properties and mode of protection, and SARPs that provide protection to other, erythromycin-related, macrolides and ketolides were also selected [73–75]. This yielded distinct consensus sequences that suggested potential drug-specific interactions within the ribosome exit tunnel [75], and contributed to the development of the 'Bottle Brush' model (**Box 4, Figure IVA**).

Especially intriguing is the recent discovery of a 61 aa protein conferring macrolide resistance that was isolated from agricultural soil bacteria [76]. Natural environments like agricultural soil are considered "hot spots" for the development of resistance when exposed to antibiotics. A macrolide-resistance conferring locus was isolated and contained several ORFs that showed no apparent linkage with any prominent resistance mechanisms. Subsequent proteomic approaches revealed that an alternative sORF was also present that had been overlooked by the initial bioinformatic predictions, which encodes for the resistance-causing poly-proline-protein PPP$^{AZI4}$. Interestingly, the first five residues of this small protein are MSWKL, and a relatively similar pentapeptide (MSWKI) had been selected from the random sequence libraries mentioned above [68, 73, 75]. The authors propose a similar 'Bottle Brush' mechanism.

**Box 4. SARPs against small molecule antibiotics.**

**1.** Protection mode against macrolides (**Figure IVA**).

The model proposed for the action of pentapeptides that increase resistance to macrolides is the "Bottle Brush" [73, 74]. Erythromycin binds the ribosome in the exit tunnel and thereby hinders nascent polypeptide elongation after several residues [77]. The small resistance peptide (MRMLT) is synthesized by the drug-bound ribosome and interacts with the drug molecule inside the exit tunnel, which eventually leads to ejection [78]. This displacement temporally cleans the ribosome for protein synthesis. Another "ribosome-quarantine" model was also proposed for other SARPs in this class [78, 79].

**2.** Protection mode against aminoglycosides (**Figure IVB**).

Selected aminoglycoside-resistance-proteins (i.e. Arp1) are highly hydrophobic and form α-helical transmembrane structures in the cell membrane. It is long known that aminoglycoside uptake relies on membrane potential and that a reduced membrane potential leads to resistance [80, 81]. Observations of synthetic Arp1 on proteoliposomes indicate that Arp1 likely acts by increasing membrane permeability to protons [31]. This causes depolarization to an extent that generates antibiotic resistance, but still maintains cell viability, as complete dissipation of the electrochemical gradient would stop bacterial growth.

**3.** Protection mode against polymyxins (**Figure IVC**).

Colistin (polymyxin E) interacts with the negatively charged phosphate groups of lipid A, a component of lipopolysaccharide [82]. The SARPs (i.e. Dcr2) interact with and activate PmrB, a membrane-localized sensor kinase that is part of the PmrAB regulon (termed BasRS in *E. coli*) [30]. Expression of ArnT, EptA and PmrR are under the control of this regulon. ArnT and EptA catalyze the modification of phosphate groups on lipid A with 4-amino-4-deoxy-L-arabinose (L-Ara4N) or phosphoethanolamine (pEtN), respectively, which lead to reduced negative charge of the cell envelope and therefore lower affinity to cationic colistin. In addition, the *29-aa* PmrR is a repressor of the phosphotransferase LpxT, which competes with EptA for a lipid A modification site. Inhibition of LpxT by PmrR therefore increases pEtN modifications.

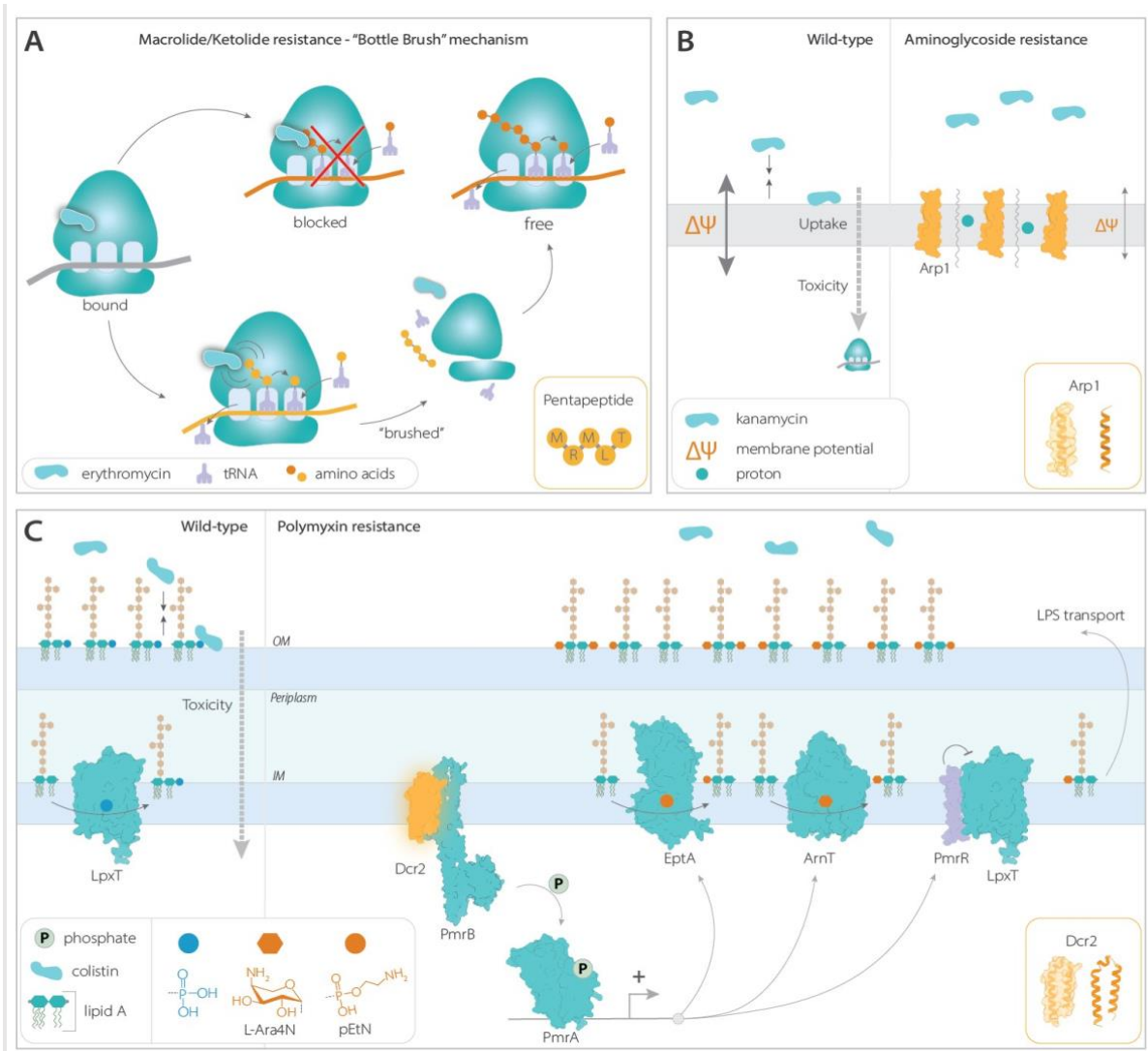

**Figure IV. Modes of action of SARPs against small-molecule antimicrobials.**

The functional role of the pentapeptide in wild-type cells remains elusive. Due to the cryptic ribosome binding site within the 23S rRNA, expression *in vivo* seems unlikely. It was speculated that perhaps stress-induced rRNA fragmentation could yield translatable fragments, or, as observed, mutations within the 23S rRNA could facilitate access to the ribosome binding site [72]. Nevertheless, it raises questions about the existence of other potentially biologically active peptides concealed within cryptic sORFs in microorganisms. The identification of PPP[AZI4] provides evidence that more, similar mechanisms exist in nature, but such sORFs can still be prone to escape detection by initial bioinformatic approaches.

## 3. SARPs evolved in the laboratory

Cells can acquire novel (resistance) genes by a number of ways, including horizontal gene transfer, rearrangements in pre-existing genetic elements, and *de novo* gene birth from previously non-coding genomic regions **(Figure 1F)**. From all resistance mechanisms, the latter is specifically of interest regarding SARPs since random occurring ORFs in non-coding genetic regions tend to be short (see also **Section 4**). SARPs selected from random sequence space on laboratory timescales provide a model for understanding such *de novo* origin **(Figure 1D)**.

A decade after the use of random sequence libraries for characterization of pentapeptides conferring resistance to macrolides and ketolides, such libraries were also used to identify SARPs against aminoglycosides in *E. coli* [31]. Notably, three unique peptides Arp1, Arp2 and Arp3 (aminoglycoside resistance peptide) increased the minimum inhibitory concentration (MIC) of different aminoglycosides up to 48-fold. These SARPs, ranging between 22 and 25 aa, are hydrophobic and localize to the cell membrane where they protect the cell by altering membrane permeability, ultimately diminishing aminoglycoside uptake **(Box 4, Figure IVB)**. This is reminiscent of some naturally evolved type I TA-system toxins, such as TisB (29 aa), which target the cytoplasmic membrane to induce a reduction in the proton motive force that can likewise increase antibiotic tolerance [83, 84].

In fact, when sequence variations of the Arps were created by site-directed mutagenesis this yielded several toxic peptides [31].

In a similar study, six SARPs that confer resistance to colistin (polymyxin E) were selected in *E. coli*, resulting in up to a 16-fold increase in MIC [30]. Colistin is currently considered a last-line therapy against multidrug-resistant gram-negative bacteria. Because of its cationic properties, colistin can directly interact with negatively charged phosphate groups in lipid A, a component of lipopolysaccharide, leading to membrane permeabilization and eventually cell lysis [82]. The selected SARPs were designated "Dcr" (*de novo* colistin resistance) and ranged from 26 to 51 aa in length. Despite lacking sequence homology, they all contain predicted transmembrane helices. Subsequent experiments unveiled an intricate mechanism involving modified lipid A headgroups, resulting in reduced colistin affinity [30] (**Box 4, Figure IVC**). Such mechanism is well-known in gram-negative organisms, but generally caused by chromosomal mutations [85]. Both these and the aminoglycoside SARPs were active in strains of *Salmonella enterica serovar Typhimurium* and *Klebsiella pneumoniae*, which are pathogens of significant clinical importance [30, 31].

To emphasize, all selected peptides were hydrophobic, which is also a feature of the immunity proteins discussed above (**Box 2**).

The relevance of findings from such libraries in understanding emergence of *de novo* resistance in clinical settings is a topic of debate [86]. Library proteins are typically overexpressed from plasmids with strong promoters, whereas *de novo* proteins tend to be expressed at lower levels before selection drives higher expression [14, 87]. Furthermore, only a few proteins were selected from millions of protein variants, sometimes providing only moderate protection. Evolutionary selected antitoxins or immunity proteins typically do not adversely affect cell growth when overexpressed [32, 88–91], similarly to some artificial selected proteins [30], but the expression of Arp1 in *E. coli* reduced cell fitness in nutrient-rich growth medium [31]. Under these conditions, wild-type cells would rapidly outcompete the Arp1-expressing cells, rendering gene fixation unlikely. However, in the presence of aminoglycoside antibiotics, Arp1 confers a strong fitness advantage and such conditional benefits could potentially evolve to become inducible through evolution of regulatory elements. It is speculated that tightly regulated beneficial toxic peptides, like TisB, have evolved in a similar manner [31]. Such small proteins could potentially also contribute to moderate resistance to antimicrobials, enabling a timeframe to develop more resistance mechanisms.

## 4. The transient small-proteome as a potential reservoir for antimicrobial resistance.

Advances in bioinformatic predictions, **ribosome profiling** and mass spectrometry reveal that many previously annotated 'intergenic' regions or non-coding RNAs are actually pervasively transcribed and translated into small proteins, in various microorganisms [14, 17, 92–94] **(Figure 1E)**. Processed peptides further add to the pool of cellular small proteins [19, 95] **(Figure 1A)**. This sparks the question – could this largely unexplored protein space form a reservoir for the generation of SARPs?

Microbes express many **noncanonical** small proteins that have not yet been annotated or characterized. On one hand this includes many small **conserved** proteins: A large-scale comparative genomics study investigated protein-coding genes less than 50 codons in metagenomic sequencing data from human microbiomes [94], and, based on evolutionary gene signatures and amino acid conservation, 4.539 small putative conserved protein families were identified of which a large proportion currently has no homologs in genome reference databases.

In addition, there is evidence that cells express (even more) noncanonical small proteins without signs of conservation [96]. In a ribosome profiling-based study in *S. cerevisiae*, over 18.000 unannotated sORFs were identified as translated with high confidence [14]. This adds ~3 times as many new ORFs to the 5400 **canonical** annotated ORFs of this yeast that have been studied for decades **(Figure 1F)**. Only a few of these ORFs are evolving under detectable **purifying selection**, whereas the majority represents evolutionary young, transient sequences of *de novo* origin. These ORFs are generally much shorter than canonical genes and lower expressed, but there is evidence that this translatome is not just translational noise but produces stable proteins that can influence various cellular processes and can provide fitness benefits under stress conditions [14]. The authors therefore suggest the terms '**conserved translatome**' and '**transient translatome**' to designate the two distinct classes of proteins [14] **(Figure 1F)**.

Even if such transient small proteins do not perform distinct biological functions, their expression may still serve a purpose. The **proto-gene** model of *de novo* gene birth suggests that such widespread translational activity in **non-genic** regions provides a reservoir of small proteins for adaptive functional innovation [87, 97] **(Figure 1F)**. Speculatively, small proteins originating from transient sORFs could modulate cell fitness under antimicrobial stress.

Given that many small bacteriocin-immunity proteins have predicted transmembrane domains, as well as the small proteins that were selected from random sequence libraries (i.e. Arps, Dcrs), it is of interest that part of the predicted *de novo* small proteins have similar hydrophobic properties [14, 94, 97, 98]. Notably, approximately one-third of all predicted conserved small proteins in the human microbiome have a predicted α-helical transmembrane domain [94]. For emerging ORFs in *S. cerevisiae*, those with fitness benefits are enriched for transmembrane helix propensity [87]. An example is the ORF *YBR196C-A* (49 codons) which emerged *de novo* from an ancestral non-genic region [87]. The hydrophobic tendency of *de novo* proteins could have a genetic basis, as intergenic regions tend to be thymine-rich and the resulting codons are enriched for hydrophobic amino acids [87, 98, 99]. Membrane localization may further provide a niche for novel proteins, shielding them from proteasome degradation, minimizing promiscuous cytoplasmic interactions, and allowing evolution of specific local interactions [87].

## 5. Concluding Remarks

The overlooked class of small proteins is now gaining increasing recognition. Small proteins are involved in various coping mechanisms to stress, including antimicrobial stress. Several small resistance proteins have been identified that protect microbes against intracellular toxins, membrane-acting protein toxins and small molecule antibiotics, either in nature or laboratory screens, as highlighted here. There is a special role for hydrophobic domains within these proteins, and the majority likely functions by modulating the activity of the toxin or other proteins by protein-protein interactions.

Recent studies show pervasive transcription and translation from 'noncoding' intergenic regions in cells. While there is evidence that some SARPs could have emerged following gene duplication events and sequence divergence (i.e. entericidinAB [32], SpoIISBC [34, 100]), or derived from segments of larger proteins [9, 36], others speculatively may have evolved *de novo* from previously non-genic regions (**Figure 1F**). Small resistance proteins could subsequently further evolve into larger multifunctional proteins by fusion to additional protein domains, as observed for intracellular TA system families, consistent with the proto-gene model of gene birth [97].

It is tempting to speculate that resistant clinical isolates of bacteria and fungi in which no resistance mechanism was identified yet may carry overlooked SARPs.

Given the rise in antimicrobial resistance to conventional small-molecule antibiotics, there is increasing interest in new antimicrobial agents, including protein toxins [101, 102]. Interestingly, natural SARPs have been reported in greater numbers against protein toxins compared to small-molecule antibiotics (which are currently the predominant anti-infective agents). It is an interesting question whether small-protein resistance could become more prevalent once protein-based pharmaceuticals were to be widely applied in clinical settings. Self-protection determinants such as SARPs can potentially be transferred from producer cells, via environmental bacteria, to clinical pathogens [103]. Immunity mimicry is a potential concern that could limit the deployment of bacteriocins in clinical practice [104]. In fact, while the bacteriocin ABP-118 (from *Lactobacillus salivarius*) protected mice during a *L. monocytogenes* infection, it was also shown that the pathogen became resistant once it was equipped with the cognate small resistance protein [105].

Some areas of these small resistance proteins deserve more attention. Several databases have been developed to organize information about protein toxins, such as BACTIBASE [106], but do not include self-immunity mechanisms. Understanding mechanisms underlying development of resistance is however important at the earliest stages of antimicrobial development. Future studies should therefore also aim to elucidate the mechanisms of protection, with special emphasis on dedicated small hydrophobic proteins against membrane-acting toxins, since many protein toxins of interest target membranes (see **Outstanding Questions**).

Continuous advances in prediction, identification, characterization and annotation of small proteins in genome reference databases will aid future studies. Gaining predictive capacity for small-protein resistance in microbial resistomes based on genome sequencing would be key to predict the sensitivity of target cells. This still represents a challenge, as small proteins may lack specific protein domains or sequence specificity for comparative genomics [30, 107].

Random sequence libraries offer a unique tool to explore presence of small resistance proteins in parts of the vast theoretical sequence space in growth-based assays, and support the concept that expression of small proteins from non-coding genetic regions could spark origination of novel small resistance genes. To the best of our knowledge, no studies have investigated the evolutionary trajectory of the transient translatome in bacteria or fungi under antimicrobial selection pressure.

Small resistance proteins also provide opportunities for biotechnology, synthetic biology and pharmacology. Specific small proteins can increase robustness of cells under stressful industrial conditions [26], and random sequence libraries could be screened for such small regulatory proteins, potentially without high metabolic cost. Resistant cells factories are also required when aiming to produce high levels of antimicrobials. In synthetic biology, fusion of an erythromycin-conferring resistance pentapeptide to a protein of interest has served as a direct transcriptional reporter and allowed selection of clones with high protein production when applying different levels of erythromycin selection pressure [108]. Such small resistance proteins may also be attractive selective markers for vector construction. As therapeutics, peptides can be used to neutralize clinically relevant toxins, and better understanding of the natural functions of SARPs could guide drug design [1, 109].

In summary, we believe it is of interest to further connect the fields of antimicrobial resistance, small proteins, cellular adaptive responses to stress, and the transient expression of noncanonical proteins to get a better understanding of this phenomenon. The space of small proteins may hold an untapped opportunity for discovery of novel resistance genes with clinical implications or biotechnological applications.

## Declaration of interests

The authors declare no competing interests.

**Outstanding Questions**

- What are the modes of protection of small hydrophobic resistance proteins that protect cells against the action of membrane-targeting toxins?

- What is the molecular basis of the difference between small proteins with a predicted transmembrane domain that are beneficial (small resistance proteins), versus those that are toxic to the cell (antimicrobial peptides)?

- Do *de novo*, transient sORFs contribute to antimicrobial resistance?

- To what extent have SARPs gone undetected in clinically relevant isolates of bacteria and fungi, and do they contribute to isolate-specific differences in antimicrobial susceptibilities?

- Given that the majority of reported natural SARPs is associated with protein-toxin resistance, can an increase in prevalence of such route of resistance be expected if protein-based antimicrobials are to become widely applied in medicine in the future? Many reports mention that resistance to membrane toxins does not arise quickly but this is often tested in controlled laboratory conditions. How would this relate to i.e. the human gut microbiome where natural antitoxin genes and horizontal gene transfer are prevalent?

- Most SARPs that have been reported are encoded as independent sORFs. To what extent can SARPs be encoded as cryptic peptides within larger proteins that get released proteolytically **(Figure 1A)** such as shown for the K2-immunity case [9]?

- How can genes encoding small resistance proteins be detected in genomic data?

- Can random sequence libraries (potentially with a bias towards hydrophobic residues) serve as a unique tool to predict, validate and characterize potential SARP mechanisms?

- How do such libraries compare to the natural available sequence space in cellular non-genic regions and rates of *de novo* gene birth?

- What are the origins and limitations of SARPs?

**Glossary:**

**Bacteriocin**: Toxic protein secreted by bacteria that can kill other, often related, microorganisms.

**Canonical**: Canonical ORFs are those that have been annotated as protein-coding genes in genome databases, the sum of which forms the canonical translatome.

**Conserved translatome**: The sum of all translation products that show signs of selection based on evolutionary gene signatures and amino acid conservation. These are mostly canonical proteins.

**Contact-dependent effector**: Toxic protein that is injected into neighboring cells upon close contact by specialized bacterial secretion systems.

**Immunity protein**: A protective protein synthesized inside a toxin-producer to prevent self-killing.

**Killer toxin**: Toxic protein secreted by yeast that kills other, often closely related species.

**Noncanonical:** Noncanonical ORFs are those marked as dubious or pseudogenes in genome databases, and includes unannotated ORFs, the sum of which forms the noncanonical translatome.

**Non-genic**: Regions in a genome that are predicted not to contain protein-coding genes are considered non-genic.

**Orphan**: In the context of a biosynthetic gene cluster, an orphan gene is a gene that is found in absence of the other biosynthetic genes it is commonly associated with.

**Proto-gene**: An ORF that has acquired all requirements for transcription and translation and has the potential to become an established gene, preserved by purifying selection.

**Purifying selection**: Randomly mutated gene variants with deleterious effects are eliminated from the population. This can result in stabilization of a sequence.

**Random sequence libraries**: A collection of random DNA sequences pooled together. Combination with transcription and translation elements and cell transformation allows for clonal production of polypeptides with random amino acid sequences.

**Resistance protein:** A protein encoded by an acquired resistance gene that protects a non-producer strain against a toxin or an antimicrobial small molecule.

**Ribosome profiling:** Also known as Ribo-Seq, a method based on deep sequencing of ribosome-protected mRNA fragments. It provides a "snapshot" of all ribosome occupation sites and therefore translated regions at a specific time point.

**Small antimicrobial resistance protein (SARP):** A term suggested herein that consolidates all small proteins and peptides of ~ 70 aa and less that confer resistance to an antimicrobial agent, including small molecules, antimicrobial peptides and protein toxins. It is an umbrella term for all small resistance-conferring proteins found within existing terminology such as "immunity proteins", "antitoxins", and other.

**Small protein**: Defined herein as polypeptides consisting of 100 aa or less. Small proteins can be derived from i.e. the expression of sORFs or from proteolytic cleavage of larger proteins.

**Toxin-antitoxin systems:** These systems are composed of two elements that are produced intracellularly by the same cell: A toxic protein and an antitoxin which inhibits the toxin.

**Transient translatome**: The sum of all proteins that are translated but lack any conserved homologs or selective signature, and are potentially evolutionary young and short-lived. These are mostly noncanonical proteins.

**Supplementary Information - Tables**

Small antimicrobial resistance proteins (SARPs) – Small proteins and peptides that confer resistance to antimicrobials.

Rianne C. Prins, Sonja Billerbeck

Molecular Microbiology, Groningen Biomolecular Sciences and Biotechnology Institute, University of Groningen, The Netherlands

**General notes and content**

**pTP:** Predicted Topology: I (in) O (out) SP (signal peptide) G (globular) - i.e. IOI indicates a predicted in-out-in topologyPredicted transmembrane regions are underscored in the amino acid sequence (DeepTMHMM, (Hallgren et al., 2022))

**pI**: The protein isoelectric point (pI) was calculated using the Expasy tool (https://web.expasy.org/compute_pi/).

**Table S1. Bacteriocins**

For bacteriocins amino acid sequences some features were highlighted; signal sequence, GG from GG-leader sequences, cysteines

Bacteriocin classification (Alvarez-Sieiro et al., 2016)

| | |
|---|---|
| Ia | lantibiotics |
| Ib | cyclized peptides |
| Ic | Sactibiotics |
| Id | linear azol(in)e-containing peptides |
| Ie | glycocins |
| If | lasso peptides |
| IIa | pediocin-like bacteriocins |
| IIb | two-peptide bacteriocins |
| IIc | leaderless bacteriocins |
| IId | non-pediocin-like, single-peptide bacteriocins |
| III | large-molecular-weight and heat-labile antimicrobial proteins |

**Table S2**. TA-systems

 - Classification is based on the mode of action of the antitoxin as described in (Singh et al., 2021)

**Table S3**. Random library SARPs
   - contains those small molecule antibiotic resistance sequences selected from random sequence libraries

**Table S4.** Yeast and others
   - contains the K2-im sequence from yeast
   - contains additionally identified sequences

# Table S1. Bacteriocins

## 1.1. Bacteriocin-Immunity proteins ≤ 70 amino acids

| | Reference | Pub med ID | Bacteri ocin | Class | Organism | Immunity gene | length (aa) | UniprotI D | pI | pTP | Amino acid sequence of immunity protein | Bacterioci n gene | amino acid sequence bacteriocin |
|---|---|---|---|---|---|---|---|---|---|---|---|---|---|
| 1 | (Skaugen et al., 1997) | 9079 878 | Lactoci n S | Ia | *Lactobacill us sake* | *lasJ* | 57 | | 9.90 | IOI | MKYSNKS<u>VMLT MLFVSVLGTVM SFSF</u>SGML<u>KAYE LILATFFLIVAVI</u> YFFKFRNEKK | *lasA* | MKTEKKVLDEL SLHASAKMGAR DVESSMNADST PVLASVAVSME LLPTASVLYSDV AGCFKYSAKHH C |
| 2 | (Heidrich et al., 1998) | 9726 851 | Epicidi n 280 (EciA) | Ia | *Staphyloco ccus epidermidi s* | *eciI* | 62 | O54219 | 9.63 | IOI | MNIY<u>LKIVITALF FSSIIFTVTYV</u>SSK NLG<u>MSLLFGLLS FIANLVYDYVM</u> GASERKSKKDN K | *eciA* | MENKKDLFDLE IKKDNMENNNE LEAQSLGPAIKA TRQVCPKATRF VTVSCKKSDCQ |
| 3 | (Aftab Uddin et al., 2021) | 3404 5548 | Homic orcin | Ia | *Staphyloco ccus hominis* | *homI* | 62 | | | | unidentified | *HomA* | MENNKNLFDLE IKKDNVENNNE LEAQSLGTAIKA TKNACPKVTRL VTVSCQKSDCQ |
| 4 | (Kabuki et al., 2011) | 2118 2576 | Thermo philin 1277 | Ia | *Streptococ cus thermophil us* | *tepI* | 52 | | 10.0 2 | IOI | MKKY<u>IPLICFLLF IVFLGITV</u>RAFLA DKT<u>LMVADGLL SIVFFISFLI</u>TRKK L | *tepA* | MMNATENQIFV ETVSDQELEMLI GGADRGWIKTL TKDCPNVISSIC AGTIITACKNCA |
| 5 | (Pag et al., 1999) | 9925 587 | Pep5 | Ia | *Staphyloco ccus epidermidi s* | *pepI* | 69 | Q57052 | 9.74 | IOI | MNIY<u>LKVILTSLF FALIIFIVTY</u>ITTK QWGT<u>SLGFSSLS FIGNFI</u>YDYSTKL | *pep5* | MKNNKNLFDLE IKKETSQNTDEL EPQTAGPAIRAS VKQCQKTLKAT |

| | | | | | | | | | | | | |
|---|---|---|---|---|---|---|---|---|---|---|---|---|
| | | | | | | | | | | SDKKYEKRINSN KKDKL | | RLFTVSCKGKN GCK |
| 6 | (Martínez-Bueno et al., 1998) | Enterocin AS-48 | Ib I | *Enterococcus faecalis* | as-48D1 | 56 | O53028 | 10.00 | IOI | MKRIDYIIIVSLL ATIVAIFLIGIDS MLGKIFLAISLGF FSSPFLKWINKLI NKN | as-48 | MVKENKFSKIFI LMALSFLGLAL FSASLQFLPIAH MAKEFGIPAAV AGTVLNVVEAG GWVTTIVSILTA VGSGGLSLLAA AGRESIKAYLK KEIKKKGKRAVI AW |
| 7 | (Kemperman et al., 2003) | Circularin A | Ib I | *Clostridium beijerinckii* | cirE | 49 | Q7WYT8 | 10.22 | IOI | MNKKKLLIYAIL FLIYIILFLTYNNS IFRIILVVSLGFLS SIISKLQIK | cirA | MFLVAGALGVQ TAAATTIVNVIL NAGTLVTVLGII ASIASGGAGTL MTIGWATFKAT VQKLAKQSMA RAIAY |
| 8 | (van Belkum et al., 2010) | Carnocyclin A | Ib I | *Carnobacterium maltaromaticum* | cclI | 60 | E1U2S0 | 9.87 | OIO | MTLLNYIFGFVM LGYGIHQYLKYQ TMYVKTNKKKY KLISLIFIVAVVCI ILGSILRLLNL | cclA | MLYELVAYGIA QGTAEKVVSLI NAGLTVGSIISIL GGVTVGLSGVF TAVKAAIAKQG IKKAIQL |
| 9 | (Gabrielsen et al., 2014) | Garvicin ML | Ib I | *Lactococcus garvieae* | garB | 55 | U2XRT5 | 10.19 | OIO | MNYNLLSITIAFI FSIHLILKNVNKK NKNWLIIIFLVSC MLIFLFGIVGILL RWF | garML | MFDLVATGMA AGVAKTIVNAV SAGMDIATALS LFSGAFTAAGGI MALIKKYAQKK LWKQLIAA |
| 10 | (Kawai et al., 2009) | Gassericin A | Ib II | *Lactobacillus gasseri* | gaaI | 53 | B3XVS0 | 11.10 | IOI | MQKLLRIIALISLI AAIISFFIFKINYIT YILIGIFIGSGFIY QIRAQGRNRK | gaaA | MVTKYGRNLGL NKVELFAIWAV LVVALLLTTANI YWIADQFGIHL ATGTARKLLDA MASGASLGTAF |

| | | | | | | | | | | | | | |
|---|---|---|---|---|---|---|---|---|---|---|---|---|---|
| | | | | | | | | | | | | | AAILGVTLPAW ALAAAGALGAT AA |
| 1 1 | (Acedo et al., 2015) | 2568 1186 | Acidoci n B | Ib II | *Lactobacill us acidophilu s* | *aciI* | 53 | A0A0C5 GQT3 | 11.1 0 | IOI | MQKLLRIIALISLI AAIISFFIFKINYIT YILIGIFIGSGFIY QIRAQGRNRK | *acdB* | MVTKYGRNLGL SKVELFAIWAV LVVALLLATANI YWIADQFGIHL ATGTARKLLDA VASGASLGTAF AAILGVTLPAW ALAAAGALGAT AA |
| 1 2 | | | Butyriv ibriocin AR10 | Ib II | *Butyrivibri o fibrisolven s* | *bviD* | 53 | Q9ZGP6 | 9.38 | IOI | MLKVTCRILSIVL GIIAIVLLGWLW MNDKREYSFIAL LIAVAGMFCLRL SRME | *bviA* | MSKKQIMSNCIS IALLIALIPNIYFI ADKMGIQLAPA WYQDIVNWVS AGGTLTTGFAII VGVTVPAWIAE AAAAFGIASA |
| 1 3 | (Borrero et al., 2018) | 2903 0449 | Plantari cyclin A | Ib II | *Lactobacill us plantarum* | *plcI* | 54 | PCL9805 1.1 | 9.51 | IOI | MKNLDMLVRVI TIILLATITAFFF KGLSTITYICAIIT VVLAFVYQLIKR HTD | *plcA* | MLSAYRSKLGL NKFEVTVLMIIS LFILLFAT**V**NIV WIAKQFGVHLT TSLTQKALDLLS AGSSLGTVAAA VLGVTLPAWAV AAAGALGGTAA |
| | | | | | | | | | | | | | |
| 1 4 | (Zheng et al., 2000) | 1080 9709 | Subtilo sin A | Id | *Bacillus subtilis* | *albB* | 53 | P71010 | 10.5 2 | IOI | MSPAQRRILLYIL SFIFVIGAVVYFV KSDYLFTLIFIAIA ILFGMRARKADS R | *sboA* | MKKAVIVENKG CATCSIGAACLV DGPIPDFEIAGA TGLFGLWG |
| 1 5 | (Duarte et al., 2018) | 2870 5677 | Hyicin 4244 | Id | *Staphyloco ccus hyicus* | *hycB* | 54 | A0A221 C8U8 | 9.99 | IOI | MSKNFLRLIIVIIF VISTVVFILKND WVFASIFGIIFIVF LLRLIQGVYNDK NQ | *hycS* | MEQGVMVSNK GCSACAIGAAC LADGPIPDFEVA GITGTFGIAS |
| 16 | (Britton et al., 2020) | 3166 7956 | Carnobacteriocin XY | IIb | *Carnobacterium piscicola* | *cbnZ* | 42 | Q46311 | 11.17 | IO | MKNFFKKNNML YRFFAVIGLIFGG WALFNIAMFIGR SIGSLF | *cbnX* | MKSVKELNVKE MQQTIGGWGW KEVVQNGQTIFS AGQKLGNMVG KIVPLPFG |
| 17 | (Proutière et al., 2023) | 3695 1576 | Gallocin A | IIb | *Streptococcus gallolyticus* | *gip* | 55 | | 8.89 | OIO | MIIKYSIIIFVNLV CYLLINKVFKAS NDERETTGKVLL ILSIVYIVVDILFN ASK | *gllA1* | MSLNKFTNFQE LDKNHLQTISG GKGNMGSAIGG CIGGVLLAAAT GPITGGGAAMI CVASGISAYL |
| 18 | (Mccormick et al., 1998) | 9835 559 | Brochocin C | IIb | *Brochothrix campestris* | *brcI* | 53 | O85758 | 9.66 | OIO | MVKTILFSVVISF VALCNFLIKKDV SSKKKLFLTGSIA VFLIIYDFLWIIFS N | *brcA* | MHKVKKLNNQ ELQQIVGGYSS KDCLKDIGKGI GAGTVAGAAG GGLAAGLGAIP GAFVGAHFGVI GGSAACIGGLL GN |
| 19 | (Ishibashi et al., 2014) | 2514 9515 | Enterocin NKR-5-3-az | IIb | *Enterococcus faecium* | *enkIaz* | 52 | A0A077 L7K7 | 9.90 | OIO | MVKALIFSAVISL IAVGNYLKKKDL PSKKKLNVTVTI AIFLIIYEFFWSIL T | *ent53a* | MQKFQKLNEQE MKQLMGGYSS KDCLKDIGKGI GAGTVAGAAG GGLAAGLGAIP GAFVGAHFGVI GGSAACIGGLL GN |
| 20 | (Flynn et al., 2002) | 1193 2444 | ABP-118 | IIb | *Lactobacillus salivarius* | *abpIM* | 55 | Q8KWH 8 | 9.05 | OI | MVFWVTVIVAY SIFISIYTSLKEDG HYKLFDEKVIEK NKKISYVRAWLL NFMWWL | *abp-118α* | MMKEFTVLTEC ELAKVDGGKR GPNCVGNFLGG LFAGAAAGVPL GPAGIVGGANL GMVGGALTCL |
| 21 | (Vera Pingitore et al., 2009) | 1959 1924 | Salivaricin CRL 1328 | IIb | *Lactobacillus salivarius* | *salIM* | 55 | | 9.05 | OI | MVFWVTVIVAY SIFISIYTSLKEDG HYKLFDEKVIEK | *salα* | MMKEFTVLTEC ELAKVDGGKR GPNCVGNFLGG LFAGAAAGVPL |
| | | | | | | | | | | | NKKISYVRAWLL NFMWWL | | GPAGIVGGANL GMVGGALT**C**L |
| 22 | (Marciset et al., 1997) | 9162 062 | Thermo philin 13 | IIb | *Streptococ cus thermophil us* | *ORF-C* | 52 | O54456 | 7.88 | OIO | MDYKS**LLSLLLFI IPLVVILIG**LRKN NQK**LIIAGVAPLI YLLCSYLLD**WIF D | *thmA* | <u>MNTITICKFDVL DAELLSTVE**GG**</u> YSGKD**C**LKDM GGYALAGAGSG ALWGAPAGGV GALPGAFVGAH VGAIAGGFA**C**M GGMIGNKFN |
| 23 | (Barrett et al., 2007) | 1741 6691 | Salivari cin P | IIb | *Lactobacill us salivarius* | *slnIM* | 44 | | | | unidentified | *sln1* | <u>MMKEFTVLTEC ELAKVD**GG**KR</u> GPN**C**VGNFLGG LFAGAAAGVPL GPAGIVGGANL GMVGGALT**C**L |
| 24 | (O'Shea et al., 2011) | 2198 4788 | Salivari cin T | IIb | *Lactobacillu* | *slnTIM1/bi m1p* | 59 | I0B596 | 9.66 | OIO | MI<u>**FWVTVIVVYS IFIS**</u>IYTSKKEDG HYKLFDEKAIKK NKKISYK**RAWIL NFIYWF**YGFY | *salTα* | <u>MMKEFTILTECE LAKVD**GG**</u>YTPK N**C**AMAVGGGM LSGAIRGGMSG TVFGVGTGNLA GAFAGAHIGLV AGGLA**C**IGGYL GSH |
| 25 | (O'Shea et al., 2011) | 2198 4788 | Salivari cin T | IIb | *Lactobacill us salivarius* | *slnTIM2* | 54 | | 9.85 | SP | MI<u>**FWVTLAVPYII FISIYT**</u>SKKEDGH YRLFDEKVIKKN KKISYVVAIVLRI SSK | *salTα* | <u>MMKEFTILTECE LAKVD**GG**</u>YTPK N**C**AMAVGGGM LSGAIRGGMSG TVFGVGTGNLA GAFAGAHIGLV AGGLA**C**IGGYL GSH |
| | | | | | | | | | | | | | |
| 26 | (Sánchez-Hidalgo et al., 2003) | 1262 0853 | Enteroc in EJ97 | IIc | *Enterococc us faecalis* | *Ej97C* | 66 | Q8KMU1 | 8.61 | OIO | MEVS<u>LIFLLGIFC VFIFFIIKLIKEKR</u> KINVGE<u>ILFIVICV CIIVLLGVFL</u>IRG DFVERPIIISPT | *ej97a* | MLAKIKAMIKK FPNPYTLAAKLT TYEINWYKQQY GRYPWERPVA |
| 2 7 | (Criado et al., 2006) | 1702 1217 | Enteroc in Q | IIc | *Enterococc us faecium* | *entQc* | 67 | Q0PHX4 | 9.63 | OIO | MLILSVIFLIIFILV ATYLIKLIKINEK KVLLSLYLIVVV FVVLFSVWFILS GNASNYKIYYGP TTI | *EntQA* | MNFLKNGIAKW MTGAELQAYK KKYGCLPWEKI SC |
| 2 8 | (Worobo et al., 1995) | 7768 812 | Divergi cin A | IId | *Carnobact erium divergens* | *dviA* | 56 | Q3SAX2 | 9.82 | IOI | MKIKWYWESLIE TLIFIIVLLVFFYR SSGFSLKNLVLG SLFYLIAIGLFNY KKINK | *dvnA* | MKKQILKGLVI VVCLSGATFFST PQQASAAAPKIT QKQKNCVNGQ LGGMLAGALG GPGGVVLGGIG GAIAGGCFN |
| 2 9 | (Franz et al., 2000) | 1074 6765 | Enteroc in B | IId | *Enterococc us faecium* | *eniB* | 58 | O85791 | 10.0 2 | IOI | MNLKKNNLEYN LCIFLAVIINLGLF IFSETILSKILLLIA IVLLVIPNFMQK KKRKNS | *entB* | MQNVKELSTKE MKQIIGGENDH RMPNELNRPNN LSKGGAKCGAA IAGGLFGIPKGP LAWAAGLANV YSKCN |
| 3 0 | (Franz et al., 2000) | 1074 6765 | Carnob acterioc in A | IId | *Carnobact erium piscicola* | *cbiA* | 56 | Q9REY7 | 9.43 | IOI | MKTNLPKEMLIL FSISIFSNLILIFLT DTIIQKVLSSLSLI ILLVVVCKEVKK NSN | *CbnA* | MNNVKELSIKE MQQVTGGDQM SDGVNYGKGSS LSKGGAKCGLG IVGGLATIPSGP LGWLAGAAGVI NSCMK |
| 3 1 | (Mathiesen et al., 2005) | 1600 0763 | Sakacin Q (SppQ) | IId | *Lactobacill us sakei* | *spiQ* | 60 | Q5ZF23 | 9.82 | IOI | VSLNKRVAYWL TQTTNLLLITILL LPAILDQEYHAIF YLLILVAIINNVV VIFFLRRQLK | *SppQ* | MQNTKELSVVE LQQILGGKRAS FGKCVVGAWG AGAAGLGAGVS GGLWGMAAGG IGGELAYMGAN GCL |
| 3 2 | (Izquierdo et al., 2009) | 1927 3675 | Enteroc in IT/Bact | IId | *Enterococc us faecium* | *entIM* | 55 | Q1T7E3 | 10.2 9 | IOI | MKNRSFAILVLS VILLDILLITLAL KTLNLVTSFVIIP | *entIT* | MKKTKLLVASL CLFSSLLAFTPS VSFSQNGGVVE |

| # | Reference | Name | abbr. | Organism | Gene | Length | UniProt | pI | Class | Immunity sequence | Gene | Sequence |
|---|---|---|---|---|---|---|---|---|---|---|---|---|
| | | eriocin 32 | | | | | | | | VIIGMSIYFIKKN KQFF | | AAAQRGYIYKK YPKGAKVPNKV KMLVNIRGKQT MRTCYLMSWT ASSRTAKYYYY I |
| 33 | (Swe et al., 2010) | Dysgal acticin | III | *Streptococcus dysgalactiae* | *dysI* | 57 | B0BLQ7 | 9.26 | IOI | MDRYEKKAFILT RLFIYISVITSFVI ENLNYKFIFYSL GTLLVIYDIFETY YNFKKK | *dysA* | MKKLKRLVISL VTSLLVISSTVP ALVYANETNNF AETQKEITTNSE ATLTNEDYTKL TSEVKTIYTNLI QYDQTKNKFYV DEDKTEQYYNY DDESIKGVYLM KDSLNDELNNN NSSNYSEIINQKI SEIDYVLQGNDI NNLIPSNTRVKR SADFSWIQRCLE EAWGYAISLVT LKGIINLFKAGK FEAAAAKLASA TAGRIAGMAAL FAFVATCGATT VS |
| 34 | (Pons et al., 2004) | Microci n L | micro cin | *Escherichi a coli* | *mclI* | 51 | B3LEN8 | 11.2 6 | IOI | MKTWQVFFIILPI SIIISLIVKQLNSS NLVQSVVSGIAI ALMISIFFNRGK | *mclC* | MREITLNEMNN VSGAGDVNWV DVGKTVATNGA GVIGGAFGAGL CGPVCAGAFAV GSSAAVAALYD AAGNSNSAKQK PEGLPPEAWNY AEGRMCNWSP NNLSDVCL |

| # | Reference | Pubmed ID | Bacteriocin | Class | Organism | Immunity gene | length | Uniprot | pI | pTP | amino acid sequence of immunity protein | Bacteriocin gene | amino acid sequence bacteriocin |
|---|---|---|---|---|---|---|---|---|---|---|---|---|---|
| 35 | (Rodríguez et al., 1998) | 9783429 | Microcin H47 | microcin | *Escherichia coli* | *mchI* | 69 | O86200 | 9.60 | IOI | MSYKKLYQLTAIFSLPLTILLVSLSSLRIVGEGNSYVDVFLSFIIFLGFIELIHGIRKILVWSGWKNGS | *mchB* | MREITESQLRYISGAGGAPATSANAAGAAAIVGALAGIPGGPLGVVVGAVSAGLTTAIGSTVGSGSASSSAGGGS |
| 36 | (Kitagawa et al., 2019) | 29980745 | rapA | | *Rhodococcus erythropolis* | *rapB* | 65 | A0A2Z6G7T1 | 12.02 | IOI | MNTQKRRQLSWRAALGIGVLYILVGLFFLSNGTTTAAVIFVILGVITALVPWVQKWWTARSAVKH | *rapA* | MIGKRVLTMAAVCAATTGVLGMGVGVASADSLPPSVCSSATGCSGNGNANPINGTNTIKDRAGCAIGLGMNAFPATWPVRVIGVGAGAYFGCN |

**1.2 Putative small immunity proteins in bacteriocin gene clusters ≤ 70 amino acids**

| # | Reference | Pubmed ID | Bacteriocin | Class | Organism | Immunity gene | length | Uniprot | pI | pTP | amino acid sequence of immunity protein | Bacteriocin gene | amino acid sequence bacteriocin |
|---|---|---|---|---|---|---|---|---|---|---|---|---|---|
| 1 | (Ito et al., 2009) | 19666732 | Gassericin A | Ib | *Lactobacillus gasseri* | *gaaC* | 60 | B3XVR8 | 9.61 | IOI | MDKKTKILFEVLYIICIIGPQFILFVTAKNNMYQLVGSFVGIVWFSYIFWYIFFKQHKKM | *gaaA* | MVTKYGRNLGLNKVELFAIWAVLVVALLLTTANIYWIADQFGIHLATGTARKLLDAMASGASLGTAFAAILGVTLPAWALAAAGALGATAA |
| 2 | (Leer et al., 1995) | 7551031 | Acidocin B | Ib | *Lactobacillus acidophilus* | *ORF1/AciC* | 60 | Q48498 | 9.61 | IOI | MDKKTKILFEVLYIICIIGPQFILFVTAKNNMYQLVGSF | *acdB* | MVTKYGRNLGLSKVELFAIWAVLVVALLLATANIYWIADQFGIHLATGTARKLLDAVASGA |
| | | | | | | | | | | IOI | VGIVWFSY IFWYIFFK QHKKM | | SLGTAFAAILGVT LPAWALAAAGAL GATAA |
| 3 | (Borrero et al., 2018) | 29030449 | Plantaric yclin A | Ib | *Lactobacillus plantarum* | plcC | 56 | PCL9804 7.1 | 9.14 | IOI | MSKFSKVS IGVIYVICT IVPAIITIFE RKLFWIGL VALGYFC YIGWFIFIK SHDNL | plcA | MLSAYRSKLGLN KFEVTVLMIISLFI LLFATVNIVWIAK QFGVHLTTSLTQ KALDLLSAGSSLG TVAAAVLGVTLP AWAVAAAGALG GTAA |
| 4 | (Xin et al., 2020) | 32487738 | Cerecyclin | Ib | *Bacillus cereus* | cycI | 58 | A0A6G5 Q8V9 | 9.84 | IOI | MNPQKSK KNSNINFS EIFALIFSIC ICIYFFGFN FKIISLSLG SLLIYFLIK FIQRKQ | cycA | MLFNVVSKLGWT GINIGTANALIGAI MTGSDIWTAISVA GIAFGGGIGTAIST IGRKAIMEMVEK VGKKKAAQW |
| 5 | (Potter et al., 2014) | 24574434 | Aureocy clicin 4185 | Ib | *Staphylococcus aureus* | aciI | 60 | X4YJ57 | 10.02 | OIO | MKTINFIF LIFLILFQI HQILKNYH LYKKTNA RKNKNLV WIMVFTLL ITCTAVLA LIFDI | AclA | MLLELTGLGIGTG MAATIINAISVGL SAATILSLSISGVAS GGAWVLAGAKQ ALKEGGKKAGIA F |
| 6 | (Xin et al., 2021) | DOI: 10.1016/j.f oodcont.20 20.107696 | Bacicycli cin XIN-1 | Ib | *Bacillus sp.* Xin1 | bacI | 60 | A0A7D5 HUK9 | 10.08 | OIO | MKIINYLF LAILFIFQI HQILKHRY AYKNTQK KYHKNMV IVFSIILVL TSISIIATY LNI | BacA | MLFELTGIGIGSG TAATIVNWIMWG MSAATILSLSISGV ASGGAWILAGAR EALKAGGGKKAAI AW |
| 7 | (Wang et al., 2021) | 33811023 | Toyoncin | IIc | *Bacillus toyonensis* | ORF7 | 62 | A0A8A5 XCE4 | 6.54 | IOI | MQKFFEAI SAIGIVGY FLGKFTSIP LIDKYTLY | toyA | MINTAWKIIKALQ KYGTKAYNVIKK GGQAMYDSFMA AKAKGWTHAAW |
| | | | | | | | | | | | FGVMLMI GVIGRFIIK VINSEEET HDSNK | | WLVEHGSTLGTF YDLLKAAGLID |
| 8 | (Liu et al., 2009) | 19202107 | Bovicin HJ50 | Ia | *Streptococcus bovis* | bovI | 52 | S5MKN7 | 10.02 | IOI | MKKYIPLI CFLLFIIFL GITVRAFL ADKTLMV ADGLLSIV FFISFLITR KKL | bovA | MMNATENQIFVE TVSDQELEMLIGG ADRGWIKTLTKD CPNVISSICAGTIIT ACKNCA |
| 9 | (Lux et al., 2007) | 17704229 | | | *Streptococcus pneumoniae* | pncM | 59 | G3EBS1 | 7.78 | IOI | MDKKRIVSTIICIVFLVVSVDNFFRDLTPLLFILNIIGLSCFSVLTYIDIKEILLNISK | | |
| 1 0 | (Lux et al., 2007) | 17704229 | | | *Streptococcus pneumoniae* | pncG | 54 | A0A4I7H216 | 9.87 | IOI | MKKKILIIFVLYLIMSIFLYLLRESAWYQLFYTIAYVIAVMIYFAINKKKGEKK | | |
| 1 1 | (Lux et al., 2007) | 17704229 | | | *Streptococcus pneumoniae* | pncB | 53 | A7XHC1 | 9.63 | OIO | MILKYSIIIAINLLSYLLTYKISKLSKNHENKIVSKILIILSIVYVIVDALLS | | |
| 1 2 | (Vaillancourt et al., 2015) | 25659110 | Suicin 3908 | Ia | *Streptococcus suis* | suiI | 51 | A0A1D8H037 | 10.02 | IOI | MKKVIPY ALLTLSMF FLFNAIRIF LESRSLYV VDVLLFIV FLVCSLKF RKQ | suiA | MNNIKPEIYVQTA TDQEITLLIGGAG SGFVKTLTKDCP GFLSNVCVNIGFI SGCKNC |
| 1 3 | (Wang et al., 2014) | 24821187 | Pericin | Ia | *Clostridium perfringens* | perI | 55 | GB EDT72684.1 | 9.58 | IOI | MKLIRIISG VVSIFFIGC AAYGYYS SKTLYLAD VILGLIAIS VFAFSFLK NSKNN | perA | MMKQLDKKSKT GIYVQVASDKEL ELLVGGAGAGFI KTLTKDCPEVVS QVCGSFFGWVSA CKNC |
| 1 4 | (Yamashita et al., 2011) | 21709077 | Bacteriocin 51 | | *Enterococcus faecium* | bacB | 57 | G1UDZ7 | 10.73 | IOI | MNKKILA VIISIIVLIT MFFIFRLIF NISLQKSIL | bacA | MKNKIIKIITCLFL FGMIVGASTPLNL FPFSARGIKVEAA SSRYNHNHRGFT |
| | | | | | | | | | | | YLIPIAFIL AIFRSIYG NKNSK | | CYKMNYYVTNK MCRDLKKNYNK LKTPAQIASFIPIG GVGTWVITNAFS SAVDAMNVFVRA ANQGKGVQLTYN AHFSNTTSYQYN DYARYVIK |
| 15 | (Georgalaki et al., 2012) | 23122510 | Macedovicin | Ia | *Streptococcus macedonius* | *mdvI* | 52 | | 10.02 | IOI | MKKYIPLI CFLLFIVFL GITVRAFL ADKTLMV ADGLLSIV FFISFLITR KKL | *mdvA* | MMNATENQIFVE TVSDQELEMLIGG ADRGWIKTLTKD CPNVISSICAGTIIT ACKNCA |
| 16 | (Vaughan et al., 2003) | 14660366 | *bronchoc in C (?)* | | *Lactobacillus sakei* | *ORF16* | 52 | Q6XVG8 | 9.90 | OIO | MVKALIFSAVISLIALGNYLKKKDLPSKK KLILTVAIAIFLIIYEFFWSSFA | | |

# Table S2. TA systems

**2.1 (Larger, canonical) antitoxins of which a (modular) smaller neutralization domain ≤ 70 amino acids has been experimentally identified**

| | Reference | Pubmed ID | TA system | Type | Organism | Antitoxin | length (aa) | Minimized neutralization domain (residues) | UniprotID | Amino acid sequence of minimized neutralization domain | Full amino acid sequence |
|---|---|---|---|---|---|---|---|---|---|---|---|
| 1 | (Sterckx et al., 2016) | 2699697 | PaaA2-ParE2 | II | Escherichia coli | PaaA2 | 63 | 13-57 | Q8XAD5 | ETIEQENSYNEWLRAKVATSLADPRPAIPHDEVERRMAERFAKMR | MNRALSPMVSEFETIEQENSYNEWLRAKVATSLADPRPAIPHDEVERRMAERFAKMRKERSKQ |
| 2 | (Jin et al., 2015) | 25622615 | VapBC4 | II | Mycobacterium tuberculosis | VapB4 | 85 | 55-85 | P9WF21 | IGELVRLGPDTTNLGEELRETLTQTTDDVRW | MSATIPARDLRNHTAEVLRRVAAGEEIEVLKDNRPVARIVPLKRRRQWLPAAEVIGELVRLGPDTTNLGEELRETLTQTTDDVRW |
| 3 | (Walling & Butler, 2016) | 27672196 | VapBC2 | II | Haemophilus influenzae | VapB2 | 77 | 46-77 | Q4QLV9 | QSWDSFFLNDQAVSDDFMNEREIAFQPEREAL | MIEASVFMTNRSQAVRLPAEVRFSEEIKKLSVRVSGSDRILSPLNQSWDSFFLNDQAVSDDFMNEREIAFQPEREAL |
| 4 | (Kamada & Hanaoka, 2005) | 16109374 | YefM-YoeB | II | Escherichia coli | YefM | 83 | 58-83 | P69346 | PANARRLMDSIDSLKSGKGTEKDIIE | MRTISYSEARQNLSATMMKAVEDHAPILITRQNGEACVLMSLEEYNSLEETAYLLRSPANARRLMDSIDSLKSGKGTEKDIIE |
| 5 | (Bernard & Couturier, 1991) | 2034222 | CcdAB | II | Escherichia coli | CcdA | 72 | 31-72 | P62552 | TMQNEARRLRAERWKAENQEGMAEVARFIEMNGSFADENRDW | MKQRITVTVDSDSYQLLKAYDVNISGLVSTTMQNEARRLRAERWKAENQEGMAEVARFIEMNGSFADENRDW |
| 6 | (De Jonge et al., 2009) | 19647513 | CcdAB | II | Escherichia coli | CcdA | 72 | 37-72 | P62552 | RRLRAERWKAENQEGMAEVARFIEMNGSFADENRDW | MKQRITVTVDSDSYQLLKAYDVNISGLVSTTMQNEARRLRAERWK |

| | | | | | | | | | | |
|---|---|---|---|---|---|---|---|---|---|---|
| | | | | | | | | | | AENQEGMAEVARFIE MNGSFADENRDW |
| 7 | of (J. Zhang et al., 2003) | 128107 11 | Ma zEF | II | *Escheric hia coli* | MazE | 82 | 38-82 | P0AE72 | VDGKLIIEPVRKEPVFTLAELVN DITPENLHENIDWGEPKDKEV W | MIHSSVKRWGNSPAV RIPATLMQALNLNIDD EVKIDLVDGKLIIEPVR KEPVFTLAELVNDITP ENLHENIDWGEPKDK EVW |
| 8 | (G. Y. Li et al., 2006) | 164135 77 | Ma zEF | II | *Escheric hia coli* | MazE | 82 | 54-77 | P0AE72 | TLAELVNDITPENLHENIDWGE PK | MIHSSVKRWGNSPAV RIPATLMQALNLNIDD EVKIDLVDGKLIIEPVR KEPVFTLAELVNDITP ENLHENIDWGEPKDK EVW |
| 9 | (Ruangpra sert et al., 2017) | 281643 93 | Din j- Yaf Q | II | *Escheric hia coli* | DinJ | 86 | 45-86 | Q47150 | KALPFDLREPNQLTIQSIKNSEA GIDVHKAKDADDLFDKLGI | MAANAFVRARIDEDL KNQAADVLAGMGLTI SDLVRITLTKVAREKA LPFDLREPNQLTIQSIK NSEAGIDVHKAKDAD DLFDKLGI |
| 1 0 | (Agarwal et al., 2010) | 200229 64 | Pe mI K | II | *Bacillus anthracis* | PemI | 95 | 50-95 | Q81VF5 | TKKRYQHESMRRGYIEMGKIN LGIASEAFLAEYEAAHTVERLV SGG | MSESSVTTEIVVRLPK QMVTELDGIGKQENK NRHELICQATQLLLRQ HKTKKRYQHESMRRG YIEMGKINLGIASEAFL AEYEAAHTVERLVSG G |
| 1 1 | (De Castroet al., 2022) | 356476 67 | Phd - Doc | II | *E.coli* Phage P1 | PhD | 73 | 52-73 | Q06253 | LDAEFASLFDTLDSTNKELVNR | MQSINFRTARGNLSEV LNNVEAGEEVEITRRG REPAVIVSKATFEAYK KAALDAEFASLFDTLD STNKELVNR |
| 1 2 | (Grabe et al., 2021) | 345568 58 | Tac AT 1 | II | *Salmonel la enterica* | TacA 1 | 88 | 53-88 | A0A455S1L 0 | NFNDEQYEEFINLLDAPVADDP VIEKLLARKPQWDV | MKSDVQLNLRAKESQ RALIDAAAEILHKSRT DFILETACQQAAEKVIL DRRVFNFNDEQYEEFI NLLDAPVADDPVIEKL LARKPQWDV |

| 13 | (Grabe et al., 2021) | 34556858 | TacAT2 | II | *Salmonella enterica* | TacA2 | 97 | 58-97 | A0A5W0DN59 | IIMADPEAYQEFLVRLDQTPSPNAALRKTMQTPAPWEQEK | MPAANSMAMKRETLNLRIKPAERDLIDRAAKARGKNRTDFVLEAARAAAEEALIEQRIIMADPEAYQEFLVRLDQTPSPNAALRKTMQTPAPWEQEK |
| 14 | (Grabe et al., 2021) | 34556858 | TacAT3 | II | *Salmonella enterica* | TacA3 | 93 | 58-93 | | YLTERDTKMIMEILDNPPAPNEKLLAAAFALPDMKK | MPQIAIESNERLSLRVSTDAKKLIVRAAAIQQTNLTDFVVSNILPVAQKIVDAAERVYLTERDTKMIMEILDNPPAPNEKLLAAAFALPDMKK |
| 15 | (Shinohara et al., 2010) | 20728434 | RelBE | II | *Pyrococcus horikoshii* | PhRelB | 67 | 50-67 | O73967 | DEDPENWIDAEELPEPED | MRMEKVGDVLKELERLKVEIQRLEAMLMPEERDEDITEEEIAELLELARDEDPENWIDAEELPEPED |

## 2.2. Contact-dependent secretion system effector-immunity proteins ≤ 80 amino acids

| | Reference | Pub med ID | Effect or syste m | Secreti on-system | *Organism* | Im mu nity | leng th (aa) | Uni prot ID | pI | Amino acid sequence of immunity | Toxin /effec tor | toxin lengt h | Uni prot ID |
|---|---|---|---|---|---|---|---|---|---|---|---|---|---|
| 1 | (Engel et al., 2012) | 2226 6942 | VbhA T | IV | *Bartonella schoenbuch ensis* | Vbh A | 62 | E6Z 0R4 | 4.5 9 | MLSEEEIEYRRRDARNALASQRLEGLEPDPQVV AQMERVVVGELETSDVIKDLMERIKREEI | VbhT | 478 | E6Z 0R3 |
| 2 | (Robb et al., 2016) | 2674 9446 | Tse2/ Tsi2 | VI | *Pseudomon as aeruginosa* | Tsi2 | 77 | PA2 703 | 4.0 5 | MNLKPQTLMVAIQCVAARTRELDAQLQNDDPQ NAAELEQLLVGYDLAADDLKNAYEQALGQYSG LPPYDRLIEEPAS | Tse2 | 158 | Q9I 0E0 |
| 3 | (González-Magaña et al., 2022) | 3633 5275 | Tse5/ Tsi5 | VI | *Pseudomon as aeruginosa* | Tsi5 | 76 | P0D TB6 | 9.4 4 (IOI ) | MPTEGRRRGVSAAMIKHYLLMTLVCIPLALLYV CLEWFFGNTWVTVGVFFGVLVVLRLGLYLYRR SKGIRDGYLDE | Tse5 | 1317 | Q9I 0F4 |
| 4 | (Mariano et al., 2019) | 317 9221 3 | Ssp6/S ip6 | VI | *Serratia marcescens* | Sip 6 | 67 | | 7.7 9 (IOI ) | MKVFSVLISRVIIGISYAVITMTLLCIAYFTLLSDS SYHIVYAIFSCIGFVLAYFIYYIAMKFHDGV | Ssp6 | 215 | |

## Table S3. Small resistance proteins isolated from random sequence libraries

| | Reference | Pubmed ID | Toxin | selected in | Resistance protein | length (aa) | pI | pTP (DeepTMHMM/ TMHMM2.0) | Amino acid sequence |
|---|---|---|---|---|---|---|---|---|---|
| 1 | (Tenson et al., 1997) (Tripathi et al., 1998) (Vimberg et al., 2004) (Tenson & Mankin, 2001) (Lovmar et al., 2006) | 9211885; 9685347; 15469510; 11587794; 16410246 | Macrolides and ketolides | Escherichia coli | many variants | 5 | | | many variants |
| 2 | (Knopp et al., 2019) | 31164464 | Aminoglycosides | Escherichia coli | Arp1 | 22 | 7.83 | SP/IO | MLLFFCFIFLLIVWLCI LAFRS |
| 3 | (Knopp et al., 2019) | 31164464 | Aminoglycosides | Escherichia coli | Arp2 | 22 | 5.28 | SP/OI | MLMFMILSLILFAIML VSAFLN |
| 4 | (Knopp et al., 2019) | 31164464 | Aminoglycosides | Escherichia coli | Arp3 | 25 | 7.82 | SP/IO | MILFTILVVVVLCLILI CTVASLTK |
| 5 | (Knopp et al., 2021) | 33411736 | Polymyxin | Escherichia coli | Dcr1 | 26 | 8.54 | SP/OI | MIVAIIIVHSLNHIMLM LLLLSLLYR |
| 6 | (Knopp et al., 2021) | 33411736 | Polymyxin | Escherichia coli | Dcr2 | 50 | 5.27 | SP/IOI | MVTTLIVLVYLIAIVL AFSFFMPTLLMTSPML TLSTLVVLLALLLLVF SN |
| 7 | (Knopp et al., 2021) | 33411736 | Polymyxin | Escherichia coli | Dcr3 | 30 | 5.27 | SP/IO | MSLFVSITFLMCIIFLCI LIMTITLTLLSL |
| 8 | (Knopp et al., 2021) | 33411736 | Polymyxin | Escherichia coli | Dcr4 | 51 | 5.28 | SP/IOI | MIILVMTIIIIVPLLTFV ILTMLSSSIVIPVVFVL LTLFLALFVIAFVTN |
| 9 | (Knopp et al., 2021) | 33411736 | Polymyxin | Escherichia coli | Dcr5 | 37 | 9.50 | SP/OI | MPIIWFITITVLLASLLL VIFLVFVVLLVIRFTFIL I |
| 10 | (Knopp et al., 2021) | 33411736 | Polymyxin | Escherichia coli | Dcr6 | 51 | 5.28 | SP/IOI | MLMFALVPLSLIFIISL LLITLVLSLPVLPVLVI LTLTMLLSAMVVVILP N |
| 11 | (Stepanov & Fox, 2007) | 17420171 | Ni$^{2+}$ | Escherichia coli | | 20 | 6.21 | G/G | MSHAYFVCNRCDSSN HSAHE |

# Table S4. Yeast systems

## 4.1 Immunity protein found in <u>yeast</u> ≤ 70 amino acids

| | Reference | Pubmed ID | Toxin | Class | *Organism* | Immunity | length (aa) | pI | Amino acid sequence of immunity protein |
|---|---|---|---|---|---|---|---|---|---|
| 1 | | | K2 | Yeast killer toxin | *Saccharomyces cerevisiae* | K2-im | 49 | 7.81 | MKETTTSLMQDELTLGEPATQARMCVR<u>LLRFFIGLTITA FIIAACIIKS</u> |

## 4.2. Other small antimicrobial resistance proteins ≤ 70 amino acids

| | Reference | Pubmed ID | Toxin | Toxin sequence | *Source* | Immunity | length (aa) | pI | Amino acid sequence of immunity protein | Notes | |
|---|---|---|---|---|---|---|---|---|---|---|---|
| 1 | (Tytler et al., 1993) | 8408070 | 18L | GIKKFLGSI WKFIKAFVG | artificial model amphipat hic helix peptides | 18A | 18 | 6.21 | DWLKAFYDKVAEK LKEAF | Recriprocal wedge-hypothesis | |
| 2 | (Shi et al., 2018) | 30558111 | ApHt_ 20 | GFWGSLWE GVKSVI | *Mesobuth us martensii* Karsch | HAP-1 | 19 | 4.08 | QKDDEEESRFFFNFI FSAE | Peptides from the chinese scorpion *Mesobuthus martensii* Karsch; HAP-1 inhibits the effect of the antimicrobial peptide ApHt_20 on *Staphylococcus aureus* | |
| 3 | (Lau et al., 2017) | 28625995 | macroli des | | | PPPAZI4 | 61 | 11.90 | MSWKLSKDPNGPL VPPRPPKLPRPPKLP RPPQPPRPPRVPQPP KPPKPPAPPGYRYT WRSK | | |